\documentstyle[12pt,prl,aps,floats]{revtex}

\begin{document}

\input{psfig}

\title{Finite Size Scaling and Running Coupling Constant in $CP^{N-1}$ models}

% AUTHOR
\author{Emanuele Manfredini}

\vspace{3mm}

% ADDRESS
\address{Dipartimento di Fisica dell'Universit\`{a}, piazza Torricelli 2, 
I-56126 Pisa, Italy \\and Istituto Nazionale di Fisica Nucleare, sezione di 
Pisa, Italy. \\E-mail: manfredi@galileo.pi.infn.it}

\date{\today}

\maketitle

% ABSTRACT
\begin{abstract}
In this work I present a numerical study of  the Finite Size Scaling (FSS) of 
a correlation length in the framework of the $CP ^{N-1}$ model by means of the
 $1/N$ expansion. This study has been thought as propedeutical to the 
application of FSS to the measure on the lattice of a new coupling constant 
$f _{x} (1/R)$, defined in terms or rectangular Wilson Loops. I give also a 
perturbative expansion of $f _{x} (1/R)$ in powers of the corresponding 
coupling constant in the $\overline{MS}$ scheme together with some preliminary
 numerical results obtained from the Polyakov ratio and I point out the 
conceptual problems that limit this approach.
\end{abstract}

\vspace{0.5truecm}

% PACS NUMBERS
\pacs{11.15.Ha; 11.15.Pg; 75.10.Jm}

% MAIN MANUSCRIPT BODY 

\section{Introduction}
\label{int}
A very important goal of a non perturbative approach to an asymptotically free
 theory such as the lattice is the determination of $\Lambda _{\overline{MS}}$
 in a physical mass unit; this requires a very precise measurement of the 
'running coupling' $\alpha _{\overline{MS}}\left( \mu \right) $ for very large
 $\mu $. In order to perform the measure by a Monte Carlo simulation on a 
lattice it is useful to work in the framework of a new renormalization scheme,
 with its own renormalized coupling constant expandable in powers of 
$\alpha _{\overline{MS}}\left( \mu \right) $. This can be done defining a 
renormalized coupling constant in terms of rectangular Wilson loops. It is 
possible then to use Finite Size Scaling techniques to reach very small 
distances without very large lattices \cite{lettera}.

In this work I want to test this program in the framework of an abelian gauge 
theory in two dimensions: the $CP ^{N-1}$ model. The choice of this specific 
model is justified by the fact that it is a renormalizable gauge theory and, 
above all, it shows both the asymptotic freedom and a confining potential 
between a particle-antiparticle pair, just as QCD does. Furthermore, 
$CP ^{N-1}$ is simpler than QCD, being an {\em abelian} gauge theory in 
{\em two} dimensions and having the possibility of an expansion in powers of 
$1/N$, that can be efficiently used to test the new renormalization scheme 
and the new approach to the problem of computing $\Lambda _{\overline{MS}}$.

This paper is organized as follows: in section \ref{cpn} I introduce the 
$CP ^{N-1}$ model both in the continuum and on the lattice. In section 
\ref{fsscl} I apply the FSS technique to the study of the correlation length, 
trying to measure the $\Lambda$-parameter in two different renormalization 
schemes. In section \ref{rccpt} I adopt the definition of the running coupling
 given in \cite{lettera}, whose perturbative expansion in powers of 
$f _{\overline{MS}} ( \mu )$ is obtained. In section \ref{polyrat} I give some
 analytical and numerical results for the new coupling constant, obtained by 
means of the Polyakov and Creutz Ratio. In section \ref{conc} I conclude with 
some considerations about the strategy outlined in the present work and I try 
to single out the ulterior difficulties that come out when one wants to apply 
the same technique to the study of the Creutz Ratio.

\section{The $CP^{N-1}$ model}
\label{cpn}

The $CP^{N-1}$ model is a generalization of the non linear $\sigma $-models.
The bare lagrangian of the continuum theory is:

\begin{equation}
\label{lag}L\left[ z\left( x\right) ,\partial _\mu z\left( x\right) ,\lambda
_\mu \left( x\right) \right] =\frac N{2f}\overline{D_\mu z\left( x\right) }
D^\mu z\left( x\right) 
\end{equation}
where $z$ is a complex $N$-vector constrained by the condition 
$\overline{z(x)}z(x)=1$ and the operator $D_\mu $ is defined as $\partial _\mu
+i\lambda _\mu $
\footnote{It can be considered as a covariant derivative associated
to a $U(1)$ local gauge invariance.}
.

As it is suggested by the mass dimension of the coupling constant $f$, the
theory is renormalizable and it shows dimensional trasmutation \cite
{cpn1,cpn2,valent}. 

Furthermore, the theory is expandable in powers of $1/N$; in this framework it
 is possible to show that the massless particles of the
`standard' perturbation theory acquire mass at the leading order in $1/N$
and they interact by a linear confining potential \cite{cpn1,cpn2,pot}.

A very good reference for the theory of spin models on the lattice is 
\cite{spin}, where we find a convenient lattice action of the $CP ^{N-1}$ 
model, the gap equation, the inverse propagators $\Delta ^{-1} _{\alpha}$ and 
$\Delta ^{-1} _{\theta}$ of the effective fields and the definition of a 
correlation length $\xi _L$ that I report here:
\begin{equation}
\label{csil}\xi _L^2=\frac 1{4\sin {}^2\frac \pi L}\left[ \frac{\widetilde{G}
_P\left( 0,0;L\right) }{\widetilde{G}_P\left( 0,1;L\right) }-1\right] 
\end{equation}
Where $G _P$ is the two-point correlation function of gauge invariant operator
 and $\widetilde{G _P}$ is its Fourier Transform. We can write 
$\widetilde{G}_P(k_1,k_2;L)$ as a function of the propagator of the effective 
field $\alpha $ in the framework of the $1/N$ expansion on a finite
lattice: 
\begin{equation}
\label{ftgp1/N}\widetilde{G}_P\left( k;L\right) =\frac 1{\beta ^2}\Delta
_{\left( \alpha \right) }^{-1}\left( k;L\right) +O\left( \frac 1N\right) 
\end{equation}
where $\Delta _{\left( \alpha \right) }^{-1}\left( k_1,k_2;L\right) $ is
given in \cite{spin}.
Using the gap equation on a finite lattice \cite{spin} I can write the 
function $\xi _L\left(
\beta \right) $ in a parametric form (with $m_L$ used as a parameter):
\begin{equation}
\label{csibeta}
\begin{array}{lcl}\xi _L^2\left( m_L\right) & = & \frac 1{4\sin {}^2\frac \pi 
L}\left[ \frac{\Delta _{\left( \alpha \right) }^{-1}\left( 0,0;L;m_L\right) }
{\Delta _{\left( \alpha \right) }^{-1}\left( 0,1;L;m_L\right) }-1\right]  \\ 
\beta\left( m_L\right) & = & \frac 1{L^2}\sum_p\frac 1{\widehat{p}^2+m_L^2}
\end{array}
\end{equation}

Let us consider now the infinite lattice limit of the correlation length so 
defined. Manipulating the formulae in \cite{spin}, it is easy to show that 
$\xi _{\infty} (\beta)$ has the following asymptotic form:
\begin{equation}
\xi _{\infty} \left( \beta \right) \simeq \frac{1}{8 \sqrt{3}} e ^{2 \pi \beta}
\label{alambdalat}
\end{equation} 
 
\section{Finite Size Scaling of the Correlation Length}
\label{fsscl}
A complete theory of the Finite Size Scaling is contained in \cite{fss,cardy},
 while an explanation of the strategy used in what follows and the definition 
of the various FSS functions can be found in \cite{luscher,car}. The FSS says 
that:
\begin{equation}
\label{fsslawcsi}\xi _L\left( \beta \right) \simeq f_P\left( \frac{ L}{\xi 
_L\left(
\beta \right) }\right) \xi _\infty \left( \beta \right) 
\end{equation}
or equivalently
\begin{equation}
\label{sigmacsi}
\frac{\xi _{L_2}\left( \beta \right) }{\xi _{L_1}\left(
\beta \right) } \simeq \sigma _\xi \left( \frac{L _2}{L _1},\frac{L _1}{\xi
_{L_1}\left( \beta \right) }\right) 
\end{equation}
I have used the expressions reported in \cite{spin} in the framework of the 
$1/N$ expansion, instead of a standard Monte Carlo simulations, in order to 
evaluate numerically the correlation length. The numerical computations have 
been performed by means of a FORTRAN code, choosing lattices of sizes varying 
from $L=20$ to $L=2000$.

\subsection{Computation of $a \Lambda _{LAT}$}
\begin{figure}[htb]
\centerline{
\psfig{figure=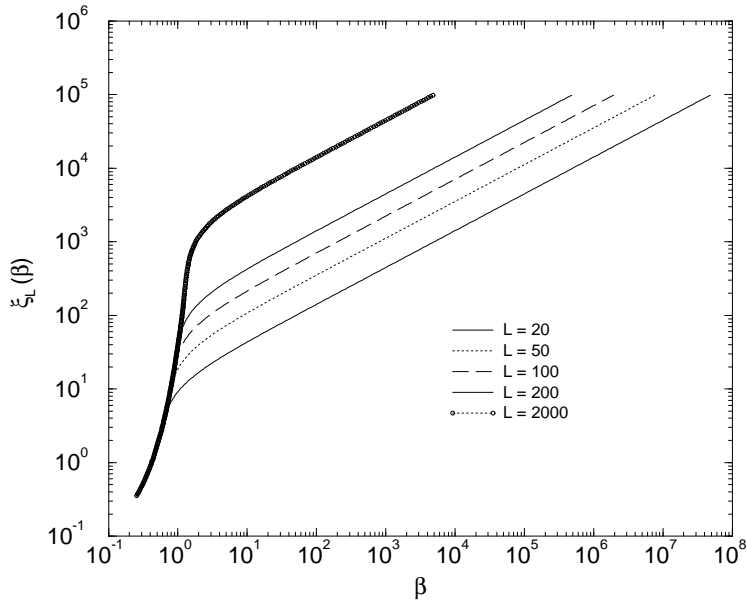,width=11truecm}}
\caption{Function $\xi _L (\beta)$ for some values of $L$.}
\label{csibetafig}
\end{figure}

\begin{figure}[htb]
\centerline{
\psfig{figure=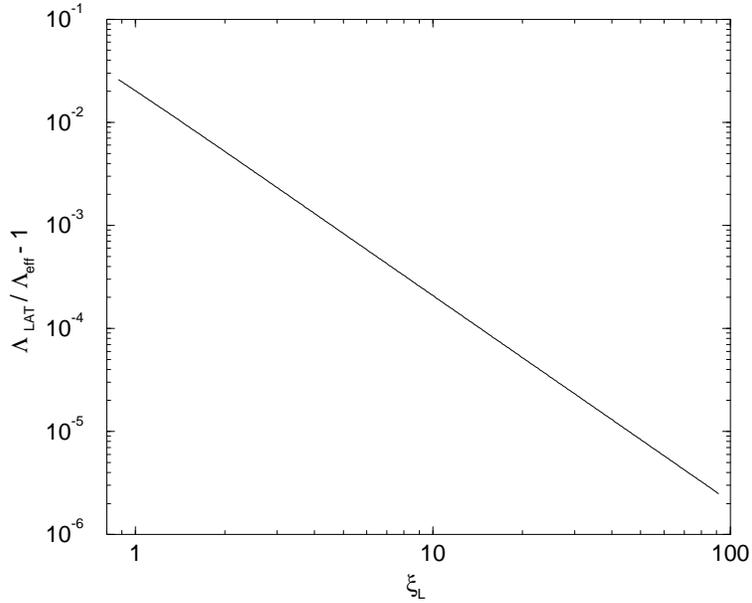,width=11truecm}}
\caption{Asymptotic scaling test.}
\label{test}
\end{figure}

The results of these computations are showed on the figure \ref{csibetafig}.
From these data I can extract a measure of the $\Lambda$-parameter in the 
lattice regularization scheme, defined by the perturbative two-loop solution
\footnote{Remember that $f = 1/ 2 \beta$.}
of the renormalization group equation for the coupling constant in the large 
$N$ limit as reported in (\ref{alambdalat}) 
\begin{equation}
\frac{\xi \left( \beta \right)}{a} = \frac1{a \Lambda _{LAT}} e ^{2 \pi \beta}
\end{equation}
The difference between $a \Lambda _{LAT}$ extracted from the lattice and the 
theoretical value $8 \sqrt{3}$ (cfr. (\ref{alambdalat})) is below $3.2 \%$.
An equivalent way to test the asymptotic scaling is to define an effective 
$\Lambda$-parameter
\footnote{The function $\xi _{\infty} (\beta)$ is exactly known from 
\cite{spin}}
:
\begin{equation}
\frac{1}{\Lambda _{eff} ( \beta )} = \xi _{\infty} ( \beta ) e ^{- 2 \pi \beta}
\end{equation}
and to plot the function 
$$
\frac{\Lambda _{LAT}}{\Lambda _{eff}} - 1
$$
versus $\xi _{\infty} (\beta)$, as showed in the figure \ref{test}.
The deviation from the asymptotic behaviour (\ref{alambdalat}) is less than 
$2.6 \times 10 ^{-6}$ for $\xi _{\infty} > 90$.   

\subsection{Computation of $\sigma _{\xi} (2, z _L)$}
\label{3.3}
\begin{figure}[htb]
\centerline{
\psfig{figure=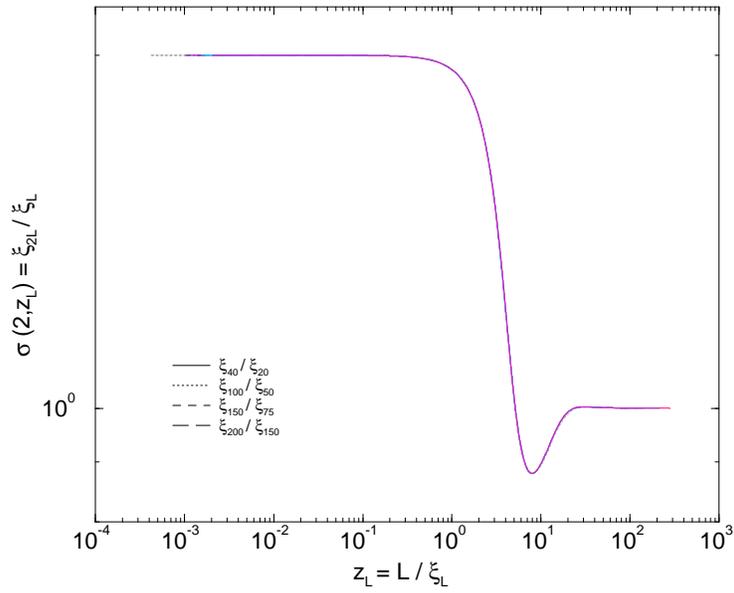,width=11truecm}}
\caption{FSS function $\sigma _{\xi} (2, z _L)$ for some values of $L$.}
\label{sigmafig}
\end{figure}

I have computed the function $\sigma _{\xi} \left( 2 , z _L \right)$ 
\footnote{$z _L = \frac{L}{\xi _L ( \beta )}$}
, according to the scheme outlined in \cite{car} and the figure \ref{sigmafig}
 shows that it has these two asymptotic behaviours:
\begin{equation}
\lim _{z _L \rightarrow \infty} \sigma _{\xi} \left( 2 , z _L \right) = 1 
~~~~~\lim _{z _L \rightarrow 0} \sigma _{\xi} \left( 2 , z _L \right) = 2
\label{limiti}
\end{equation}
that reflect these simple physical considerations:
\begin{enumerate}
\item In the limit $z _L \rightarrow \infty$ the actual size $L$ of the box is
 not very important because the correlation length $\xi _L (\beta)$ of the 
interaction is very small, compared to L; we are very close to the infinite 
volume limit.
\item In the limit $z _L \rightarrow 0$, on the contrary, the correlation 
length $\xi _L (\beta)$ is much larger than $L$, that becomes the actual scale
 of the interaction: the observables are very sensible to what happens on the 
border of the box; that is, $\xi _L (\beta)$, having the same dimension as 
$L$, is directly proportional to it.
\end{enumerate}

If the FSS is correct, the function (\ref{sigmacsi}) should depend only on the
ratio $\frac L{\xi _L}$, and not on $\xi _L$ and $L$ separately. I can
check if this is the case by verifying that the several curves obtained on the
different pairs of lattices approximately superimpose. The results are
showed on the figure \ref{sigmafig}. 
The biggest relative violation of the FSS is about $5.4\times 10^{-3}$ for 
$z_L=4.635...$

\subsection{Reconstruction of $f _{\xi} ( z _L )$}
Now, in order to go on with the program outlined in \cite{car}, I need to know
 the function $\sigma
_\xi \left( 2,z_L\right) $ in every point of a sufficiently large interval;
I have two possibilities:
\begin{enumerate}
\item interpolate the points I have obtained numerically using a polynomial of
 n-th degree, with n to be chosen.
\item find a suitable function that fits the points at our disposal very well.
 The theory \cite{Neuberger} tells us that $\sigma _{\xi} (2, z _L) 
\rightarrow 1$ exponentially fast as $z _L \rightarrow \infty$. Then I can try
 to use a fit function of the form
\footnote{$s$ is a scale factor to be chosen. An analysis of the tipical mass
 unit of the theory with $N = \infty$ suggests $s = 2 \sqrt{6}$ as the best 
choice.}
\begin{equation}
\label{fit}\sigma _\xi ^{(fit)}\left( 2,z_L\right) =1+\sum_{n=1}^{N}a_n\exp
\left( -nz_L/s\right) 
\end{equation}
with the constraint that $\lim _{z _L \rightarrow 0} \sigma _{\xi} \left( 2 , z _L \right) = 2$.

\end{enumerate}

\begin{figure}[htb]
\centerline{
\psfig{figure=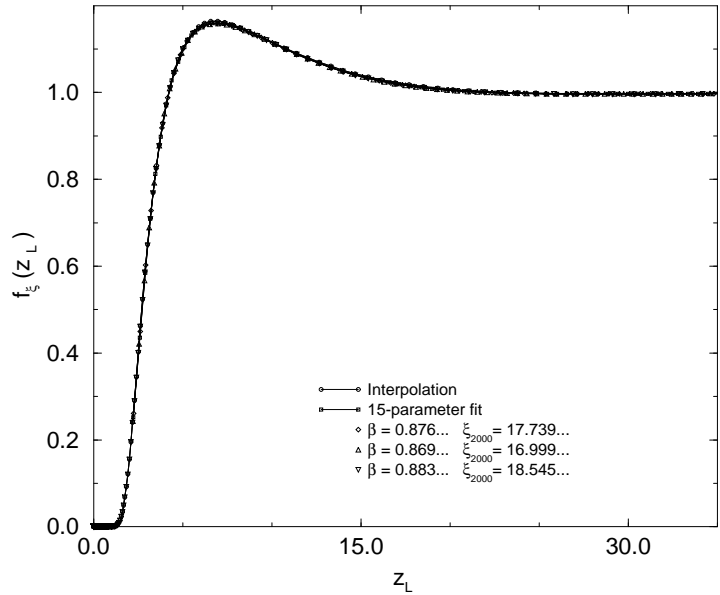,width=11truecm}}
\caption{FSS function $f _{\xi} ( z _L) $ computed in three different ways: 
a.Interpolation of $\sigma _{\xi}$. b.$15-$paramters fit of $\sigma _{\xi}$. 
c.Direct calculation of $f_{\xi}$ as $\frac{\xi_L}{\xi_{2000}}$.}
\label{effefig}
\end{figure}

There are several systematic effects intrinsic in both procedures; I have made
 some checks in order to choose the values of the parameters that minimize 
them. Anyway, all `spurious' systematic errors can be made smaller than the 
FSS violation, which in turn constitutes the real physical limit of our 
procedures. The figure \ref{effefig} shows the results of the calculation.

\subsection{Perturbative expansion of $\sigma _\xi$ and $f_\xi $}
\label{pertpar}
A perturbative calculation on a finite lattice, along the line sketched in 
\cite{hase} gives the following result for the FSS function $f_\xi$:
\begin{equation}
\label{fcsipert}f_\xi \left( z_L\right) =R\left( 4\pi \right)
^{2/N}e^{-d}z_L^{-1+4/N}e^{-4\pi /z_L^2}\left( 1+O\left( z_L^2\right)
\right) 
\end{equation}
with $R=\left( \xi _\infty \Lambda _L\right) ^{-1}=8\sqrt{3}$ and 
$d=0.76077028681...$ In the large $N$ limit:
\begin{equation}
\label{fcsipertNinf}f_\xi \left( z_L\right) =Re^{-d}z_L^{-1}e^{-4\pi/z_L^2}
\left( 1+O\left( z_L^2\right) \right) 
\end{equation}
and
\begin{equation}\label{sigmapert}\sigma _{pert}\left( 2,z_L\right) =2\left( 
1-\frac{\ln 2}{8\pi }z_L^2 - \frac{ \ln ^2 2}{64 \pi ^2} + O\left( z_L^6
\right) \right) 
\end{equation}
I checked the range of validity of this perturbative result and I found good 
agreement for all $z _L < 1$.

We can do this perturbative test on the function $f_\xi $ directly. If we look
 at the shape of the function $\sigma $$_\xi $ we realize that the number of 
points in which it is necessary to know $\sigma _\xi$ itself in order to 
reconstruct $f _\xi$ is a decreasing function of $z_L$; this implies that a 
systematic error on the numerical computation of $\sigma $$_\xi $ will 
propagate more and more when $z_L$ becomes smaller and smaller. This is 
exactly what I find if I compare the numerically reconstructed functions 
$f _{\xi} ^{(fit)}$ and $f _{\xi} ^{(int)}$ with the exact perturbative result
 (\ref{fcsipertNinf}).
It is worth noticing that I can explain this effect quantitatively quite well 
invoking the FSS violation studied at the end of the paragraph \ref{3.3}.

\subsection{An asymptotic scaling test}
\begin{figure}[htb]
\centerline{
\psfig{figure=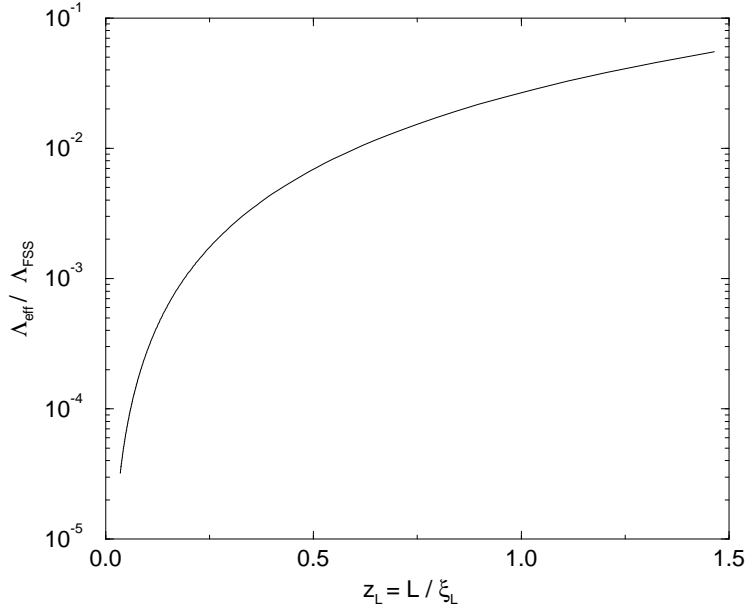,width=11truecm}}
\caption{Asymptotic scaling test for the FSS function $f _{\xi} (z _L)$}
\label{fssast}
\end{figure}

It is worth noticing that in the large $N$ limit the FSS function $f _{\xi}
 (z _L)$ is exactly known in terms of the two functions \cite{Pelissetto}:
$$
F _1 (z) = \frac{\pi}{z} - \log \pi + \frac{\log 2}{2} + \gamma _E +
$$
$$
+ 2 \pi \sum _{n=1} ^{\infty} \left( \frac{1}{\sqrt{4 \pi ^2 n ^2 + z ^2}} - 
\frac{1}{2 \pi n}  \right) +
$$
\begin{equation}
+ 2 \pi \sum _{n = -\infty} ^{\infty} \frac{1}{\sqrt{4 \pi ^2 n^2 + z^2}}
 \frac{1}{\exp \left( \sqrt{4 \pi ^2 n ^2 + z ^2} \right) - 1}
\end{equation}
\begin{equation}
F _2 (z) = \sum _{n=-\infty} ^{\infty} \frac{1}{1-4 n ^2} \frac{\coth \left( 
\frac{\sqrt{4 \pi ^2 n ^2 + z ^2}}{2}  \right)}{\sqrt{4 \pi ^2 n ^2 + z ^2}}
\end{equation}
so we can obtain the function $f _{\xi} (z _L)$ in a parametric form:
\begin{equation}
\label{fxizl}
\begin{array}{lcl}
f _{\xi} (x) & = & \frac{8 \sqrt{3}}{z _L (x)} \exp \left[ -F _1 \left(\frac{1}
{x} \right) \right] \\
z _L (x) & = & \frac{2 \pi}{\sqrt{-1 - \pi x \frac{F' _1 \left( \frac{1}{x} 
\right)}{F _2 \left( \frac{1}{x} \right)}}}
\end{array}
\end{equation}
As already seen at the beginning of this section, the asymptotic behaviour of 
this expression is given by (\ref{fcsipertNinf}); in order to test the speed 
of approach to the asymptotic scaling I define from equation 
(\ref{fcsipertNinf}) the `asymptotic function': 
\begin{equation}
f _{\xi} ^{as} (z _L) = \Lambda _{FSS} \frac{e ^{-\frac{4 \pi}{z _L ^2}}}{z _L}
\end{equation}
where $\Lambda _{FSS} = R e ^{-d}$ and $d = 0.76077028681...$.
I define also an effective $\Lambda _{FSS}$-parameter:
\begin{equation}
\Lambda _{FSS} ^{eff} (z _L) = f _{\xi} (z _L) z _L e ^{\frac{4 \pi}{z _L ^2}}
\end{equation} 
and I plot in the figure \ref{fssast} the function 
\begin{equation}
\frac{\Lambda _{FSS} ^{eff}}{\Lambda _{FSS}} - 1
\end{equation}
that shows the deviation from the asymptotic scaling; this is less than $3.5
 \times 10 ^{-5}$ for $z _L < 0.036$.

\section{The `Running Coupling' from Perturbation Theory}
\label{rccpt}
Following the procedure described in \cite{lettera}, let us consider a 
rectangular Wilson loop $(R \times T)$
\begin{equation}
  W _{ \Gamma } ( R , T ) = \left\langle e ^{i \oint _{\Gamma} A _{\mu} (t) 
d t _{\mu} } \right\rangle
\label{wl}
\end{equation}
It is possible to give a definition of an interaction force $\chi _T (R)$ in 
which a partial derivative with respect to $T$ substitutes the $\lim _{T 
\rightarrow \infty}$ of the standard definition \cite{bw,wilson}:
\begin{equation}
\chi (R,T) = \frac{\partial ^{2} \ln W(R,T)}{\partial R \partial T}.
\label{chi}
\end{equation}
The perturbative expansion, in the continuum theory, of the interaction force 
$\chi _T (R) $ so defined is:
\begin{equation}
\chi ^{\overline{MS}} _{T} (R) =\frac{N-1}{2\pi }I_1(x)\frac{t_{\overline{MS}}
 (\mu) ^2}{R^2} \left\{ 1+2b_0 \left[ \ln(\mu R)+a(x) \right] 
t_{\overline{MS}} (\mu) +O(t_{\overline{MS}} (\mu) ^2) \right\}
\label{chims}
\end{equation}
where
\begin{equation}
I_n(x)=\int \frac{d^2q}{(2\pi )^2}q^2(\ln q)^n\frac{\sin q_1}{q_1}\frac{\sin
(q_0x)}{q_0} 
\label{in}
\end{equation}
\begin{equation}
a(x)=1-\frac{I_2(x)}{2I_1(x)} 
\label{ax}
\end{equation}
$x = \frac{T}{R}$, $t _{\overline{MS}} (\mu) $ is the renormalized coupling 
constant in the modified minimal subtraction scheme and $b _0$ is the first 
coefficient of the perturbative $\beta$ function of the model.

A direct computation of the integrals $I _1 ( x )$ and $I _2 ( x )$ gives:
\begin{equation}
I _{1} (x) = f(x) + \frac{1}{x^2}f \left( \frac{1}{x} \right)
\end{equation}
with
\begin{equation}
f(x) = - \frac{1}{2 \pi} \left( \arctan x + \frac{x}{1 + x ^2} \right)
\label{a.1}
\end{equation}
and
$$ 
I _{2} (x) = \frac{\gamma _{E} -1}{2} \left( 1 + \frac{1}{x ^2} \right) - 
\frac{\ln x}{2 x ^2} - 
$$
$$
- \frac{1}{\pi} \left\{ B(x) + \frac{1}{x ^2} \arctan x \ln x + \frac{1}
{1+x ^2} \left[ \frac{\pi}{2} + \frac{1}{2} \ln (1+x ^2) \left( x + \frac{1}
{x} \right) \right] + \right.
$$
$$
+ \left. \int _{0} ^{1} \frac{du}{u} \left[ \frac{1}{x^2} \arctan \left( 
\frac{ux}{1+x ^2 (1-u)} \right) + \arctan \left( \frac{ux}{1+x ^2 -u} \right)
 \right] \right\} +
$$
\begin{equation}
+ \frac{x}{\pi} \int _0 ^1 dt \, t ^2 \ln \left( \frac{1}{t ^2} -1 \right) 
\left[ \frac{1}{(1+ x ^2 t ^2 ) ^2} + \frac{1}{(x ^2 + t ^2 ) ^2} \right] 
\end{equation}
with
\begin{equation}
B(x) = g(x) + \frac{1}{x ^2} g (\frac{1}{x})
\end{equation}
and
\begin{equation}
g(x) = (\gamma _{E} - 1) \arctan \frac{1}{x} - \gamma _{E} \frac{x}{1+x ^2} -
 \arctan \frac{1}{x} \ln \frac{x}{1+x ^2}
\end{equation}
These functions have the following $x \rightarrow \infty$ limit:
\begin{equation}
\lim _{x \rightarrow \infty} I _1 ( x ) = - \frac{1}{4} ~~~~~ \lim _{x 
\rightarrow \infty} I _2 ( x ) = \frac{\gamma _E - 1}{2} 
\end{equation}
Now it is possible to compute the function (\ref{ax}) numerically. The 
results are reported in the table \ref{taba1}.
\begin{table}[htb]
\caption{Numerical value of the function $a(x)$ for some values of $x$.}
\begin{tabular}{cccc}
$x$     &       $a(x)$ & $x$     &       $a(x)$ \\ \tableline
$1$     &       $-0.289942...$ & $4$     &       $-0.091605...$ \\
$5/4$   &       $-0.450157...$ & $5$     &       $0.007810...$ \\
$4/3$   &       $-0.472398...$ & $10$    &       $0.238997...$ \\
$3/2$   &       $-0.488345...$ & $50$    &       $0.478908...$ \\
$5/3$   &       $-0.479687...$ & $100$   &       $0.520289...$ \\
$7/4$   &       $-0.469696...$ & $200$   &       $0.544656...$ \\
$2$     &       $-0.427594...$ & $300$   &       $0.553868...$ \\
$3$     &       $-0.234557...$ & $\infty$        &       $0.577216...$ \\
\end{tabular}
\label{taba1}
\end{table}

I can now define a running coupling constant $t _x ( 1/R )$:
\begin{equation}
\chi (R,T) = \frac{N-1}{2 \pi} I _1 (x) \frac{t_{x}(1/R)^{2}}{R^{2}}~~~~~~~~~
x=\frac{T}{R} 
\label{runcc}
\end{equation}
and a large N-rescaled coupling constant
$f _x (1/R) = N t _x (1/R) / 2$
\begin{equation}
\chi (R,T) = \frac{2}{\pi} \frac{N-1}{N ^2} I _1 (x) \frac{f _{x}(1/R)^{2}}
{R^{2}}
\label{fruncc}
\end{equation}

The definition (\ref{runcc}) and the perturbative result (\ref{chims}) allow 
to establish a link between $t _x \left( \frac{1}{R} \right)$ and 
$t _{\overline{MS}} \left( \mu \right)$:
\begin{equation}
t _x \left( \frac{1}{R} \right) ^2 = t _{\overline{MS}} \left( \mu \right) ^2
 \left\{ 1 + 2 b _0 \left[ a(x) + \ln \mu R \right] t _{\overline{MS}} \left(
 \mu \right) + O \left( t _{\overline{MS}} \left( \mu \right) ^2 \right) 
\right\}
\end{equation}
where $a(x)$ determines the connection between the $\Lambda$-parameters in the
 two renormalization schemes:
\begin{equation}
a(x) = \ln \frac{\Lambda _x}{\Lambda _{\overline{MS}}}
\label{lambdaratio}
\end{equation}

\section{Analytical and Numerical results from the Polyakov Ratio}
\label{polyrat}
The Wilson line or Polyakov loop is strictly related to the static 
quark-antiquark potential \cite{rothe}. I can take the interaction potential
 in the continuum theory for the case $x \rightarrow \infty$ from \cite{pot}:
\begin{equation}
V(R)=\frac{6\pi }Nm_0^2R+\frac{2c_L}Nm_0-\frac 2N\int_0^\infty \cos
(kR) \left[ \Delta _{(\lambda )}(k)-\frac{12\pi m_0^2}{k^2} \right] \frac{dk}
{2\pi } 
\label{potnapp}
\end{equation}
where
\begin{equation}
\Delta _{(\lambda)} (k) = \frac{2 \pi}{Y(k) \ln \frac{Y(k) + 1}{Y(k) - 1} - 2}
 ~~~ Y(k) = \sqrt{1 + \frac{4 m _0 ^2}{k ^2}}
\end{equation}
After having rotated the integration contour in the complex plane and other 
algebraic manipulations I obtain
\begin{equation}
\label{vr}
V(R)=\frac{6\pi }Nm_0^2R+\frac{2c_L}Nm_0 + \frac {2 \pi}{N} \int _{2 m _0} 
^{\infty} e ^{-x R} \frac{Y}{(\pi Y) ^2 + \left( Y \ln \frac{1+Y}{1-Y} -2 
\right) ^2}
\end{equation}
from which I can extract the interaction force
\begin{equation}
\label{fr}
N F(R) \xi ^2 = \pi + \frac{\pi}{3} \int _2 ^{\infty} dt e ^{- \frac{tR}{\xi
 \sqrt{6}}} \frac{tY}{(\pi Y) ^2 + \left( Y \ln \frac{1+Y}{1-Y} -2 \right) ^2}
 = \Phi (r = \frac{R}{\xi})
\end{equation}
where in (\ref{vr}) and (\ref{fr}) $Y(t) = \sqrt{1 - \frac{4}{t ^2}}$. 

The formula (\ref{fr}) can be used to compute numerically the reference
 continuum quantity to be compared with the lattice results; it can also be 
expanded in the regime $r \rightarrow 0$ (perturbative or scaling region) 
obtaining the scaling behaviour of the interaction force:
\begin{equation}
\Phi (r) = \frac{\pi}{2} \frac{1}{r ^2 \ln ^2 \frac{\sqrt{6}}{r}} + 
\frac{\pi \gamma _E}{r ^2 \ln ^3 \frac{\sqrt{6}}{r}} + O(\frac{1}{r ^2 
\ln ^4 \frac{\sqrt{6}}{r}})
\label{conintfor}
\end{equation}
This result can be rewritten with the same precision as
\begin{equation}
\Phi (r) = \frac{\pi}{2} \frac{1}{r ^2 \ln ^2 \frac{\sqrt{6}}{ e ^{\gamma _E}
 r}} \left( 1 + O(\frac{1}{r ^2 \ln ^4 \frac{\sqrt{6}}{r}}) \right)
\label{scalingforce}
\end{equation}
It is possible to extract from (\ref{scalingforce}) the running coupling 
constant in the scheme outlined in the previous paragraph in the limit $x 
\rightarrow \infty$ 
\begin{equation}
f _{\infty} (r) = \frac{\pi}{\ln \frac{\sqrt{6}}{e ^{\gamma _E} r}}
\end{equation}
This expression is in perfect agreement with the prediction of the 1-loop 
perturbative renormalization group
\begin{equation}
\Phi (r) = \frac{\pi}{2} \frac{1}{r ^2 \ln ^2 \frac{1}{\Lambda _{\infty} R}}
\label{pertforce}
\end{equation}
if the $\Lambda$-parameter is chosen consistently with the renormalization 
scheme introduced before
\footnote{See (\ref{lambdaratio})}
:
\begin{equation}
\Lambda _{\infty} R = e ^{a(\infty)} \Lambda _{\overline{MS}} R = e ^{\gamma 
_E} r / \sqrt{6}
\label{lambda}
\end{equation}

The difference between the exact continuum curve and the asymptotic one is 
below $5 \%$ for $r < 0.25$. One could then measure the interaction force  on
 a finite lattice by means of the so-called Polyakov Ratio, for several values
 of $r _L < 0.25$, keeping $z _L$ fixed, and then fit the values with equation
 (\ref{pertforce}) in order to extract the $\Lambda$-parameter. It is so 
possible to understand how fast the measures on the lattice approach the 
continuum limit (\ref{lambda}) when the lattice spacing goes to 0.

The Polyakov Ratio on a finite lattice with periodic boundary conditions is 
given by
\begin{equation}
\chi _P (R) = \frac{1}{L} \ln \frac{W(R,L)}{W(R-1,L)} = - \frac{1}{2N} 
\frac{1}{L} \sum _k \frac{\sin \frac{k _{1} (2R-1) \pi}{L}}{\sin \frac{k _{1}
 \pi}{L}} \widehat{k _1} ^{2} \Delta _{(\theta)} (k _1,0)
\label{latpoly}
\end{equation}
in the leading order in the $1/N$ expansion, where $\Delta _{(\theta)} (k)$ is
 the lattice propagator of the gauge field $\theta$ reported in \cite{spin}.

\begin{figure}[htb]
\centerline{
\psfig{figure=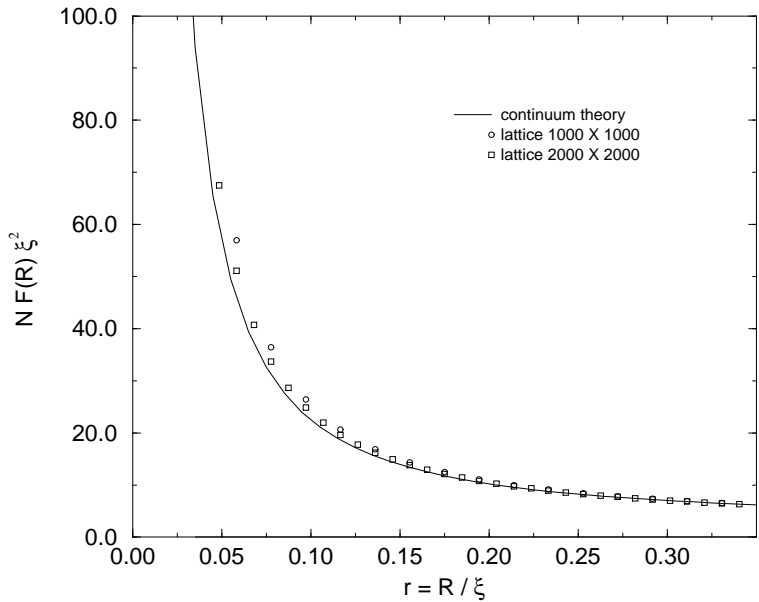,width=11truecm}}
\caption{Systematic effect $\propto a/R$ on the value of the Polyakov ratio on
 a finite lattices. The continuum line is the continuum interaction force 
\ref{conintfor}}
\label{sist}
\end{figure}
A numerical calculation of the Polyakov Ratio has been performed on two 
lattices ($L = 1000$ and $L=2000$) and for two $\xi _L$, keeping $z _L = 
19.455...$ fixed. The results are showed on the figure \ref{sist}, where we 
can see the presence of a systematic effect $\propto a/R$ for $r=R/ \xi _L <
 0.25$; when $r > 0.25$ other effects due to periodicity start to play a much
 more important role \cite{spin,finite}.

\section{Conclusions and Outlook}
\label{conc}
The study of the correlation length has clarified how to use the FSS technique
 to estrapolate the finite volume measures to infinite volume results. In this
 framework it could be interesting to apply the same technique to a basic
 observable: the Creutz Ratio derived from rectangular Wilson loops. Even if 
this seems to be just a mere step-by-step execution of the program already 
outlined in \cite{car} for the correlation length, there are some points that
 should be underlined:
\begin{enumerate}
\item the presence of the additional scale $R$ implies that the FSS function
 depends on two variables, instead of only one.
\item a systematic effect $\propto \left( \frac{a}{R} \right) ^2$ (analogous
 to that found in the case of the Polyakov Ratio) is present in the Wilson 
loop case, too.
\item in the special case of an abelian lattice gauge theory defined on a 
torus, as our version of the lattice $CP ^{N-1}$ model is, there are some 
kinematical effects that have to be considered in order to obtain significant 
results \cite{finite}.
\item the special shape of the FSS function $\sigma _{\xi} \left( 2 , z _L 
\right)$ imposes another limitation on the region where the FSS technique can 
 be applied efficiently: let us define
\begin{equation}
\Phi ( \beta , R , x , L ) = \chi ( \beta , R , xR, L )
\end{equation}
$\Phi$ obeys to a FSS law similar to (\ref{fsslawcsi}) seen in the section 
\ref{fsscl} for the correlation length:
\begin{equation}
\Phi \left( x,\beta ,R,L\right) =g_\Phi \left( x,\frac R{\xi _L(\beta
)},\frac L{\xi _L(\beta )}\right) \Phi \left( x,\beta ,R,\infty \right)  
\label{4.4}
\end{equation}
The counterpart of the function $\sigma _{\xi}$ is 
\begin{equation}
s _{a,b} (z _L , r _L) = \lim _{\beta \rightarrow \infty} \frac{\Phi (\beta, b
 R , x , a L )}{\Phi ( \beta , R , x , L )} 
\end{equation}
where
\begin{equation}
z _L = \frac{L}{\xi _{L} (\beta)} ~~~ r _L = \frac{R}{\xi _{L} (\beta)}
\end{equation}
The limit $\beta \rightarrow \infty$ means simply that we are in the scaling 
region.
If the FSS law (\ref{4.4}) is valid and choosing $a = 2$, $b = 1$
\begin{equation}
s _{2,1} (z _L , r _L ) = \frac{g _{\Phi} (x, z' _L , r' _L)}{g _{\Phi} (x, z 
_L ,r _L )} = \frac{\Phi (\beta , R, x , 2L )}{\Phi (\beta , R, x, L)}
\end{equation}
with
\begin{equation}
z' _L = \frac{2 z _L}{\sigma _{\xi} \left( 2 , z _L \right) } ~~~ r' _L = 
\frac{r _L }{\sigma _{\xi} \left( 2 , z _L \right) }
\label{4.5}
\end{equation}
Once I have measured the function $s _{2,1} (z _L , r _L )$ in a large enough 
region of the plane $(z _L , r _L)$, I can try to reconstruct the function $g 
_{\Phi} (x, z _L ,r _L )$ by:
\begin{equation}
g _{\Phi} (x, z _L ,r _L ) \simeq \prod _{j=1} ^{n} \frac{1}{s _{2,1} 
(z ^{(j)} _L , r ^{(j)} _L)} g _{\Phi} (x, z ^{(n)} _L ,r ^{(n)} _L )
\end{equation}
with
\begin{equation}
z ^{(j)} _L = \frac{2 z ^{(j-1)} _L}{\sigma \left( 2 , z ^{(j-1)} _L \right) }
 ~~~ r ^{(j)} _L = \frac{r ^{(j-1)} _L }{\sigma \left( 2 , z ^{(j-1)} _L 
\right) }
\label{80}
\end{equation}

A severe limitation to this computational scheme is represented by the
 practical impossibility to reach very small values of $L/ \xi _L (\beta)$. 
From figure \ref{sigmafig} we see that $\sigma _{\xi} \simeq 2$ for $L/ \xi _L
 (\beta) < 1$. When we try to compute $g _{\Phi}$ for a $L/ \xi _L (\beta)$
 quite small, $\sigma _{\xi} (2, z_L ^{(j)})$ in (\ref{80}) stays very close 
to $2$ even for $\overline{\j}$ very large; now, for $\j < \overline{\j}$, 
while $r _L$ is reduced to a half at each step, $z _L$ stays about constant 
(remember (\ref{4.5})). I can do two things, in order to lessen $r _L$: I can
 lessen $R$, and, once I have reached  the lowest limit for $R$, since $a$ is
 finite, I must increase $\xi _L (\beta)$, and this means that I have to 
increase $L$ too, to keep $z _L$ constant. In conclusion, if I want to compute
 the function $g _{\Phi} (x, z _L ,r _L )$ for a very small $z _L$ I need to 
simulate on very large lattices, or, in other words, a upper limit on the size
 $L$ of the lattice implies a lower limit on $z _L$, as far as the measure of 
$g _{\Phi} (x, z _L ,r _L )$ concerns.

\end{enumerate}

These considerations leads to the conclusion that the application of the FSS
 technique to the Wilson loop needs a careful study of all the systematic 
effects. Furthermore, the problem outlined in the last point above seems to 
limit the efficiency of the method in the region where also the traditional 
techniques fail. The crucial point is represented by the exact form of $\sigma
 _{\xi} ( 2 , z _L )$ when $z _L \rightarrow 0$. Once I have fixed the largest
 dimension $L _{MAX}$ of a usable lattice, the lowest limit $z _L ^{(min)}$ at
 which it is possible to compute the function $g _{\Phi} ( x , z _L , r _L )$ 
is different from $0$ and it depends on the specific definition of the 
correlation length chosen. The only thing we can do is to find a correlation
 length that minimizes $z _L ^{(min)}$.

\section{Aknowledgements}
I would like to thank P. ROSSI, E. VICARI and A. PELISSETTO for their useful
 comments and suggestions.  

% REFERENCES

\end{document}